\documentclass[]{spie}  

\usepackage{amsmath,amsfonts,amssymb}
\usepackage{graphicx}
\usepackage[colorlinks=true, allcolors=blue]{hyperref}
\usepackage[numbers]{natbib}

\title{BlackCAT CubeSat: \\
A Soft X-ray Sky Monitor, Transient Finder, and Burst Detector for
High-energy and Multimessenger Astrophysics} 

\author[a]{Tanmoy Chattopadhyay}
\author[a]{Abraham D. Falcone}
\author[a]{David N. Burrows}
\author[a]{Derek B. Fox}
\author[b]{David Palmer}

\affil[a]{Pennsylvania State University, University Park, PA 16802, USA}
\affil[b]{Los Alamos National Laboratory, Los Alamos, NM 87544, USA}

\authorinfo{Further author information: (Send correspondence to T. Chattopadhyay)\\T.C.: E-mail: txc344@psu.edu, Telephone: 1 814 852 9733\\}

\begin{document} 
\maketitle

\begin{abstract}
Here we present the conceptual design of a wide field imager onboard a 6U 
class CubeSat 
platform for the study of GRB prompt and afterglow emission and detection 
of electromagnetic counterparts of gravitational waves in soft X-rays. 
The planned instrument configuration consists of an array of X-ray Hybrid CMOS 
detectors (HCD), chosen for their soft-X-ray response, flexible and rapid readout rate, and low power, which 
makes these detectors well suited for detecting bright transients on a CubeSat platform. The wide 
field imager is realized by a 2D coded mask. We will give an overview of the 
instrument design and the scientific requirements of the proposed mission.
\end{abstract}

\keywords{X-ray Hybrid CMOS detector, CubeSat, Wide field imager, Coded mask,
X-ray transients}

\section{INTRODUCTION}
\label{sec:intro}  

The recent detection of GW 170817 / GRB 170817A by LIGO, Virgo, and Fermi -- 
followed by afterglow detections across the electromagnetic spectrum -- has 
started a new era in multimessenger astronomy. The joint set of
observations has conclusively linked binary neutron star (BNS) mergers with short-duration gamma-ray bursts (e.g. \citep{fox05}), revealed the event's host galaxy NGC 4993 and merger
site \citep{coulter17}, and provided new and valuable constraints on the neutron star (NS) equation of state \citep{radice18}.
Numerical simulations suggest bright electromagnetic (EM) counterparts to NS+BH
mergers will also be observed, in which case these events will provide additional insights into
strong-field gravity, relativistic outflows, and the NS equation of state.
The possibility of EM counterparts from binary black hole (BBH) mergers, which can yield additional physical and astrophysical insights into these events, cannot be excluded.
A global network of advanced gravitational wave (GW) facilities operating at design sensitivity, as we anticipate in five years, 
will detect these events at a 
rate of about one event per week. 

To detect such EM counterparts of gravitational waves, wide-field 
monitoring missions are necessary, with a requirement of rapidly locating these events with sub-arcminute accuracy and passing the information on to the wider astronomy community for further study. Five years down the line,
when gravitational wave detectors across the globe are expected to 
function at their highest sensitivity, the present high-energy sky 
monitoring facilities will be $\sim$10 years or more beyond their design lifetimes (2009 for Swift; 2013 for Fermi), and may have ceased operations entirely. 
On the other hand, existing or 
planned wide-field monitors are sensitive in hard X-rays. Wide-field
monitors sensitive in soft X-rays are expected to provide much better
detection probability and significantly better localization accuracy. 
In this context, we have proposed a soft X-ray wide-field imager for a CubeSat mission. CubeSats are inexpensive and will serve as pathfinder missions for future satellite missions with wide-field imaging
instruments.

The proposed design, BlackCAT\footnote{`Black' refers to black hole related astrophysics and `CAT' stands for Coded Aperture Telescope}, is based on 
coded mask imaging with an array of 4 X-ray Hybrid CMOS detectors (HCDs). HCDs
provide faster readout rate and better radiation hardness compared to CCDs,
and most importantly, these are low power devices, which is important for a CubeSat mission. For this mission, we plan to use an advanced version of HCD detectors, Speedster EXDs, which utilize event driven readout that makes these detectors an order of magnitude faster compared to the H$x$RGs. We plan to procure a standard 6U CubeSat 
platform from Clyde Space for the mission. The standard spacecraft 
meets the requirements of power, telemetry, and attitude control needed
for the mission. 
The spacecraft will point in a direction that is 
approximately anti-Sun, thus simplifying 
pointing and thermal considerations while maximizing
dark sky coverage, which in turn will facilitate ground-based follow-up of discovered GRBs.
An overview of the mission in given in Table \ref{mission}.
\begin{table}[ht]
\caption{Mission overview}
\label{tab:mission}
\begin{center}
\begin{tabular}{|l|l|} 
\hline
\rule[-1ex]{0pt}{3.5ex}  Instrument & Soft X-ray coded-mask telescope with hybrid CMOS detectors  \\
\hline
\rule[-1ex]{0pt}{3.5ex}  Spacecraft & Standard 6U CubeSat (from Clyde Space)   \\
\hline
\rule[-1ex]{0pt}{3.5ex}  Orbit & Sun-synchronous low Earth orbit   \\
\hline
\rule[-1ex]{0pt}{3.5ex}  Science & Detection of EM counterparts of gravitational waves, high redshift GRBs, transients, monitoring   \\
\hline
\end{tabular}
\end{center}
\label{mission}
\end{table}

The primary science goals of BlackCAT are as follows, 
\begin{itemize}
\item detection and localization of the EM counterparts of gravitational wave 
events and communication of this information to other facilities for detailed multiwavelength study 
of these events, 
\item discovery of GRBs at high redshifts and study of their timing and 
spectroscopic properties,
\item monitoring of known black hole X-ray binaries and detection of their 
state changes, and 
\item discovery of new black hole sources in their outbursts. 
\end{itemize}    
BlackCAT would also provide the opportunity to evaluate the Speedster detectors in space and raise the TRL level of this particular technology It should be noted that another form of X-ray hybrid CMOS detector is now at TRL-9 \citep{chattopadhyay18_HCDoverview}, while the more novel detectors chosen for this mission have been ground-characterized and are being prepared for opportunities to be flight-proven.

The design of the mission has been submitted to NASA 
and is currently under review. In Sec. \ref{science},
we  give an overview of the main science goals of the mission, followed
by a description of the instrument design and spacecraft in 
Sec. \ref{blackcat}. In Sec. \ref{plan}, we discuss the future plans
related to the realization of the instrument and mission. 

\section{Science case}\label{science}

The BlackCAT mission is designed to address multiple promising science cases, 
which are briefly described here. 


\subsection{Gamma Ray Bursts}

\begin{figure} [ht]
   \begin{center}
   \includegraphics[scale=0.35]{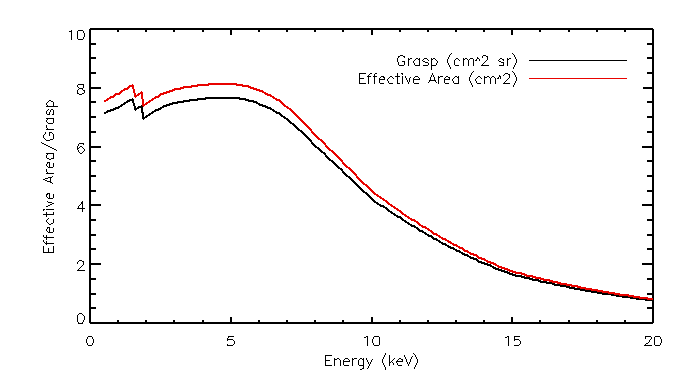}
\includegraphics[scale=0.35]{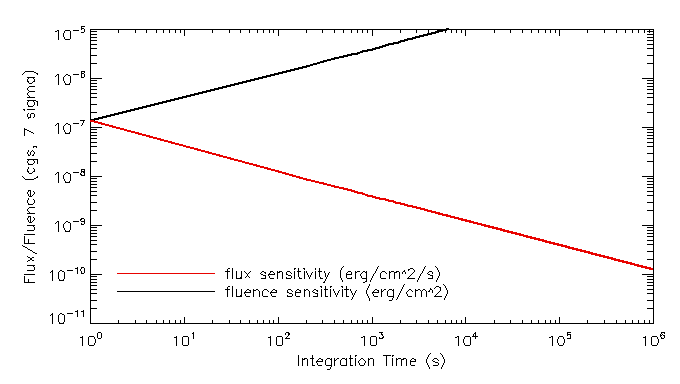}\\
\includegraphics[scale=0.35]{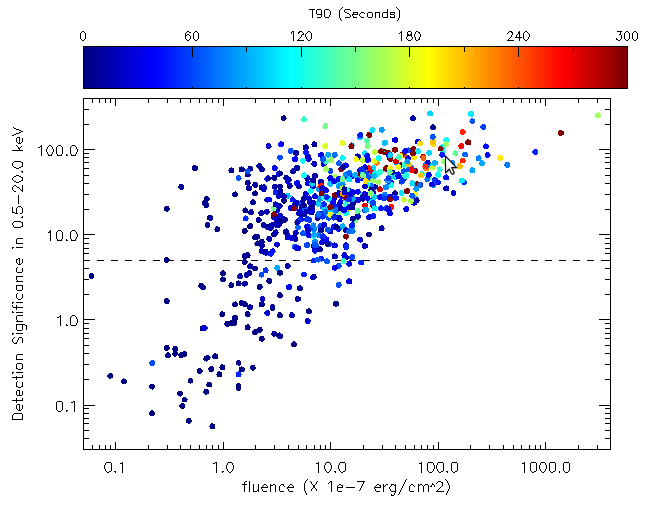}
\includegraphics[scale=0.46]{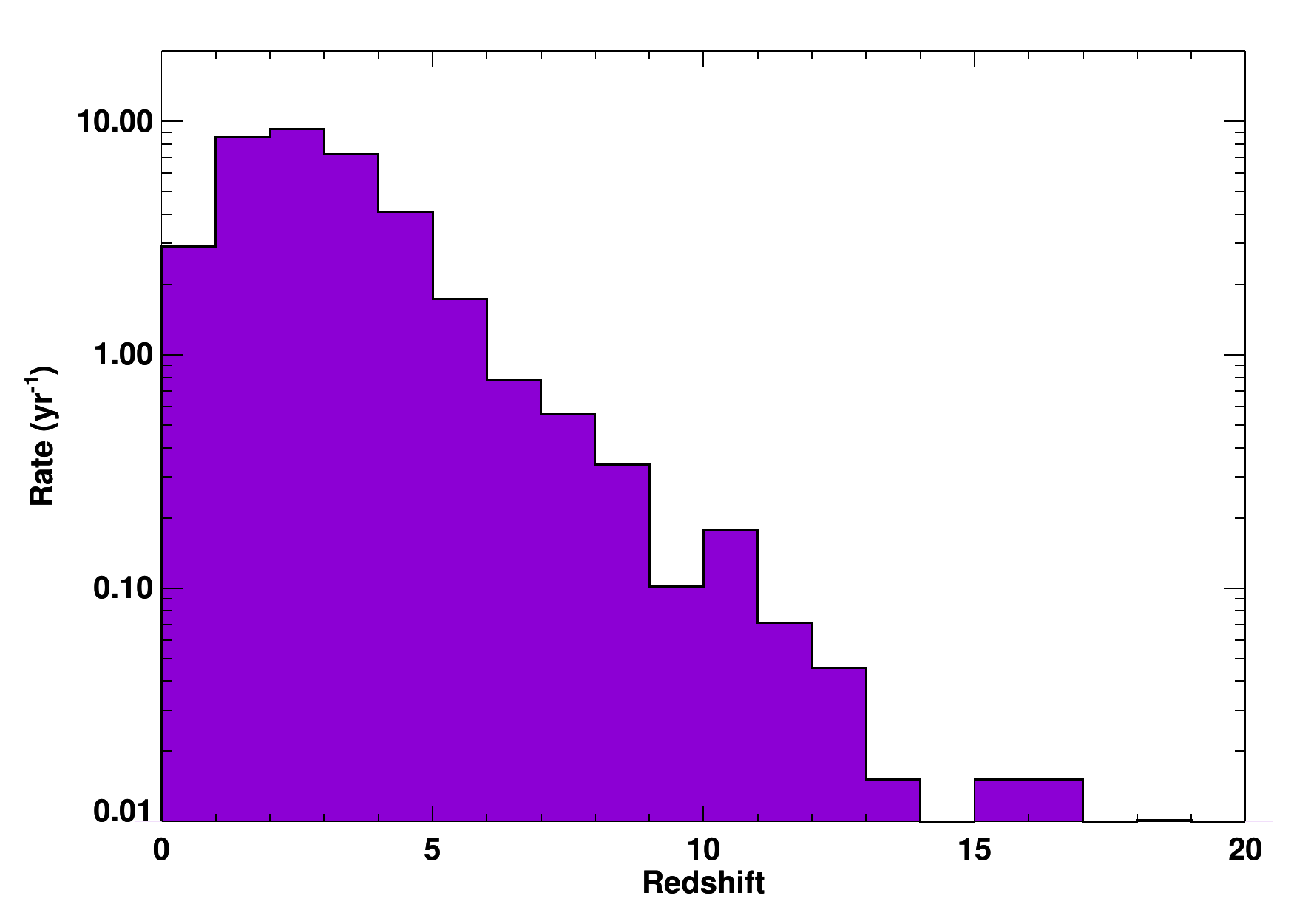}
   \end{center}
   \caption[example] 
   { \label{grb} 
Top: Effective area as a function of energy (left) and sensitivity as a function of integration time (right) for the proposed BlackCAT 
configuration. Bottom left: Distribution of GRB fluence (vertical scale) and T$_{90}$ (color scale).  The 
BlackCAT 5$\sigma$ detection
threshold is indicated by the dashed horizontal line. Bottom right: Redshift distribution of GRBs detectable with 
BlackCAT (right).}
   \end{figure}

One of the primary science goals of BlackCAT is to discover new high-redshift GRBs and study their timing and spectroscopic properties. 
Fig. \ref{grb} shows the simulated BlackCAT effective area and response 
to a range of likely gamma ray burst (GRB) redshift distribution functions. 
BlackCAT is expected to detect and provide prompt, sub-arcminute 
positions for 36 gamma ray bursts per year in the soft X-ray band, with an inherent selection bias for the highest redshift GRBs, in contrast to the higher energy bands of Swift-BAT and Fermi. BlackCAT would detect $\sim$0.8 bright short-duration bursts per year, $\sim$4.0 GRBs per year from z$>$5, and $\sim$0.8 GRBs per year from z$>$8. All detected bursts will be deep in the night-time sky (within $\sim$60$^\circ$ of anti-Sun),
perfectly positioned for prompt and extended follow-up observations by ground-based optical/NIR observatories and satellite telescopes, allowing detailed studies of the burst afterglows.
By discovering the brightest high-redshift GRB afterglows in the night 
sky and reporting their positions to observers in real-time, BlackCAT will enable high-quality spectroscopic observations from premier facilities. These spectra will be used to map out the history of the cosmic re-ionization, measure the escape fraction of ionizing radiation from high-redshift star-forming regions, and explore processes of metal 
enrichment in the early Universe.


BlackCAT's high-sensitivity characterization of the prompt X-ray spectra of 
GRBs from all redshifts will yield useful insights into the emission 
mechanism and surroundings of GRBs, which are known to have a broad peak 
in their $\nu$-F$_\nu$ energy spectra. BlackCAT will measure burst spectra down 
to $\sim$0.5 keV, testing for the presence of low-energy thermal components or spectral breaks. 
Current observations cannot distinguish between these two scenarios, 
but the questions will be easily addressed by BlackCAT's medium 
resolution 0.5-20 keV spectra of every detected burst. GRB jets
are thought to pierce their progenitor's stellar envelope and exploding SN, so that detection of
prompt emission at lower energies enables measurement of local absorption and its evolution with time. The prompt low-energy spectrum can also be searched 
for absorption lines, with any detection amounting to an exciting discovery 
and enabling an independent redshift measurement.

\subsection{Multi-messenger astronomy}
\label{sec:title}


Another primary science goal of BlackCAT is to detect and locate the 
electromagnetic counterparts of gravitational wave events. This is possible
due to the large field of view (FOV) of the instrument and its high sensitivity in soft X-rays. 
In 2022, the GW signals of coalescing binary neutron stars (BNS),
binary black holes (BBH), and neutron star + black hole (NS+BH) mergers will be detected at
rates greater than one per week by the advanced GW detectors of the Laser 
Interferometer Gravitational Wave Observatory (LIGO) in Washington and Louisiana; the Virgo detector in Italy; and
the KAGRA detector in Japan. The joint detection of 
GW 170817 / GRB 170817A by LIGO,
Virgo, and Fermi – followed by kilonova / afterglow detections across the electromagnetic (EM)
spectrum, from X-rays to radio wavelengths \citep{abott17} – 
has ushered in a new era
where multimessenger observations provide key insights into the physics 
and astrophysics of
these events. 
BlackCAT can play a vital role in discovering and locating the EM 
counterpart of such rare events. From simulation results, we find that 
GRB 170817A would have registered as a 
4.1$\sigma$ point source. Hence,
while not itself registering as a triggered burst ($>$5$\sigma$ threshold), 
it would have been readily recovered in ground analyses prompted by the 
GW observation. Moreover, a mere 50\% increase in
high-energy flux would have produced a $>$5$\sigma$ detection and 
burst trigger, resulting in a sub-arcminute position and rapid electronic 
notification to the astronomical community.

\subsection{Detection of Black hole Transients}

Due to its large FOV coupled with high sensitivity in 0.5-20 keV, BlackCAT
is expected to discover new black hole transient events and monitor the 
known sources to detect changes in their spectroscopic states.
The majority of  X-ray black hole binary (BHB) systems spend most of their time 
 in a low emission undetectable quiescent state,
and therefore there are expected to be a large undiscovered population of 
these objects. The discovery rate is 2-3 per year. 
High energy X-ray and gamma-ray emission from
these objects is currently the only means of discovery, and will likely 
remain the primary means even with future observatory development. Wide
FOV Instruments like BlackCAT are therefore essential to
monitor and detect BHBs in outburst to provide triggers for the next 
generation of observatories.
In addition to detecting transients as they rise from quiescence 
in the Low/Hard state (LHS), BlackCAT's sensitivity range of 0.5-20 keV will make it 
uniquely sensitive to detecting BHBs in the High/Soft state (HSS), where the
typical temperatures are $\sim$1 keV and the majority (99.8\%) of this thermal 
emission is below 10 keV. With typical outbursts lasting months to years, 
we expect most bright ($>$10$^{-9}$erg/s/cm$^2$) BHBs to be detected while in 
outburst. Determination of current state and changes 
of state 
of transient and persistent BHBs in this critical soft X-ray band 
will be key to providing triggering information to the
next generation of facilities dedicated to the study of accretion processes. 
\begin{figure} [ht]
   \begin{center}
   \includegraphics[scale=0.5]{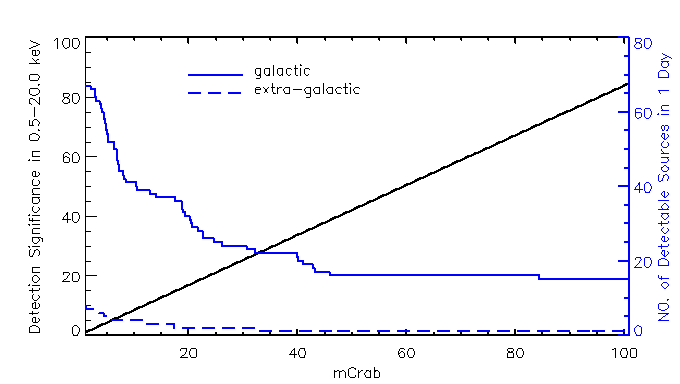}
   \end{center}
   \caption[example]
   { \label{monitoring}
Daily monitoring sensitivity of BlackCAT (black solid line) assuming 50\%
of duty cycle. The detection thresholds for galactic and extra-galactic sources in 1 day are 
shown with blue-solid and blue-dashed lines, respectively.
}   
\end{figure}
Individual sources will be monitored for up to 100 days/year.  
%

BlackCAT will monitor the high states of blazars as well. The BlackCAT 
soft X-ray energy range
is ideal for monitoring blazar flares, since many of them produce 
copious synchrotron emission
that peaks in this energy band. 
X-ray data is required for 
interpretation of blazar emission spectra measured at other wavelengths, 
such as those from TeV imaging air Cherenkov telescopes (VERITAS, HESS, 
Magic; and in the future, CTA) and from the Fermi gamma-ray space
telescope. Since pointed instruments like Swift-XRT, Suzaku, Chandra, 
and XMM can not monitor a significant fraction of the sky, monitoring 
instruments like BlackCAT are needed for the studies of cosmic ray 
acceleration and blazar emission mechanisms from these objects. There
are currently $>$50 known TeV blazars, and each of these is capable of 
producing flares that would trigger BlackCAT and lead to broad community 
multi-wavelength coverage that would otherwise
be unobtainable. 

From simulation study, we found that BlackCAT will detect BHBs down to a 
level of $\sim$6 mCrab in daily averages (5$\sigma$), with
an average of 3 BHBs in the BlackCAT FOV per day (see Fig. \ref{monitoring}). 
Outside of the flaring regime, we estimate that BlackCAT 
will detect $\sim$54 galactic sources and $\sim$5 extra-galactic sources during 
any given 1-day co-add (5$\sigma$ limit), based on
sources drawn from RXTE-ASM catalog (see Fig. \ref{monitoring}). Given that 
ASM was only sensitive above 2 keV, whereas BlackCAT is sensitive down 
to 0.5 keV, we expect these numbers to improve.

\section{BlackCAT design}\label{blackcat}

\subsection{Instrument overview}
\begin{figure} [ht]
   \begin{center}
 \includegraphics[scale=0.6]{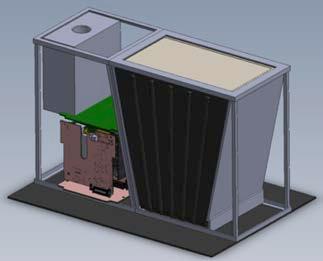}
   \end{center}
   \caption[example]
   { \label{DM}
Mechanical design of BlackCAT.}   
\end{figure}

BlackCAT is an array of soft X-ray detectors with coded mask imaging. 
A detector module (DM) is mounted to the Payload Support Structure
(PSS). The DM consists of a truncated inverted pyramid with 
four detectors in the
base, a coded aperture mask to provide imaging capability, and side 
walls designed to absorb off-axis X-rays and to support the mask 
(see Fig. \ref{DM}). 
The DM will have 3-point mounting to the payload support structure, using
Ti flexures to minimize thermal conduction and flexing. The side walls 
support the coded aperture mask above the detector mosaic and will be 
fabricated from Si Carbide trusses and Al walls to provide stable optical 
performance over the expected on-orbit temperature range (with focal 
length variations $<$100 $\mu$m). The 3.5 mm wall thickness will provide
sufficient X-ray shielding from off-axis X-rays (stray
light).

\begin{figure} [ht]
   \begin{center}
\includegraphics[scale=0.5]{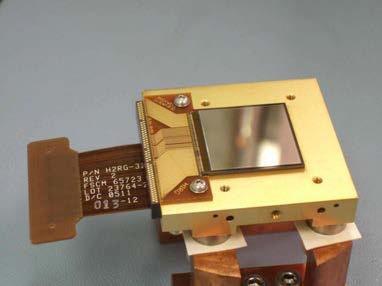}
   \end{center}
   \caption[example]
   { \label{h1rg}
Photo of H1RG-161 with deposited Al optical blocking filter. The Speedster-EXD 
flight detector in BlackCAT will be similarly sized to an H1RG.
}
   \end{figure}

\begin{figure} [ht]
   \begin{center}
   \includegraphics[scale=0.3]{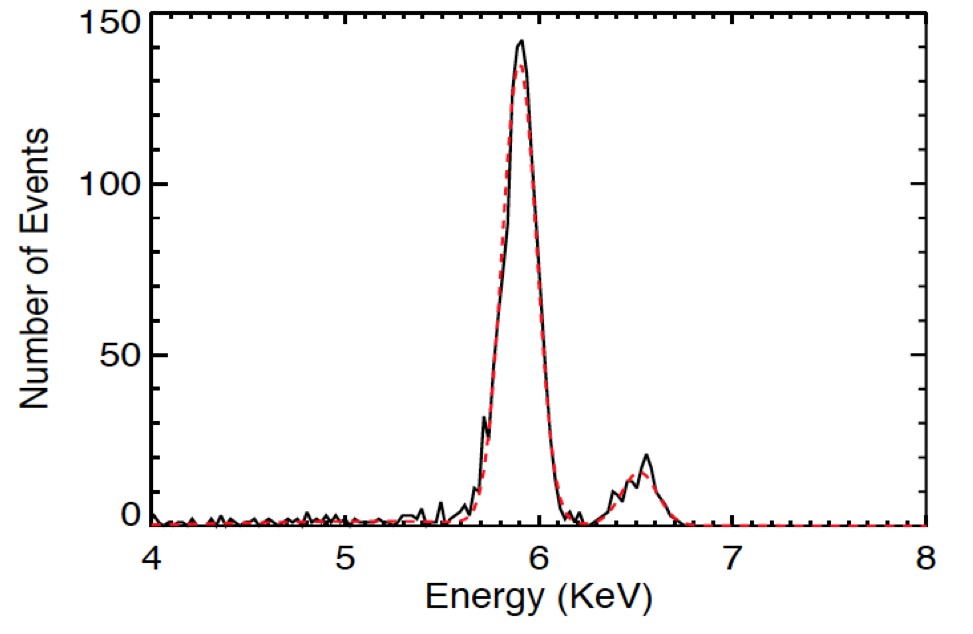}
   \end{center}
   \caption[example]
   { \label{spectra}
Spectrum of monochromatic X-ray lines collected with one of the Speedster EXD
proto-type detector (Mn K$_\alpha$ (5.9 keV) and Mn K$_\beta$ (6.4 keV)). The measured resolution at 6 keV is around 3.3 \% which exceeds the BlackCAT requirements.
}   
\end{figure}

{\bf SPEEDSTER-EXD:} BlackCAT uses a new event-driven detector design 
(Speedster-EXD) based on Teledyne HyViSI \citep{bai08}, which is a line of hybrid CMOS detectors
made of a silicon photo-absorbing pixel array connected to a CMOS
readout multiplexer via indium bump-bonds.
The Speedster-EXD is an advanced design 
utilizing a Capacitive Transimpedence Amplifier (CTIA), in-pixel CDS, 
and event recognition circuitry to provide event-driven sparse readout. 
The CTIA eliminates interpixel capacitive crosstalk completely (Griffith et al.
2016). The in-pixel CDS reduces 1/f noise and cancels the CTIA reset 
KTC noise. Event-driven readout is accomplished using a comparator that 
samples the CDS output and triggers on X-ray
events. This provides timing resolution improvements by several orders 
of magnitude and a significant improvement in high count-rate applications. The input cell architecture also supports
multiple readout nodes. The detectors are selected for their extremely 
low power requirements (compared to CCDs), their radiation hardness, 
their rapid readout, and the opportunity to flight test 
them on a CubeSat mission. Fig. \ref{h1rg} shows an image of a Teledyne
H1RG detector. The proposed flight version of the SPEEDSTER detector will be similar in size to an H1RG detector. Each
detector is 2.2 cm $\times$ 2.2 cm in size, with 550 $\times$ 550 pixels 
(40 $\mu$m pixel size). There will be 
4 such detectors in BlackCAT as mentioned earlier, with each having thickness
of 100 $\mu$m. 
On-board source detection would use 320 $\mu$m $\times$ 320 $\mu$m
“superpixels,” based on 8 x 8 binning of the physical pixels to speed and 
simplify processing. Devices with these features have been calibrated at 
PSU at X-ray wavelengths \citep{griffith16,falcone07,bongiorno09,falcone12,prieskorn13,hull17,chattopadhyay18_HCDoverview}
and meet the requirements of this mission. Fig. \ref{spectra} shows an 
energy spectrum for Mn K$\alpha$ (5.9 keV) and Mn K$\beta$
(6.4 keV) for a prototype of Speedster-EXD with 64 x 64 pixels.

{\bf Coded Mask: }The coded aperture mask will be supported by a ribbed frame.
Baseline design parameters are given in Table \ref{mask}. The mask is made 
of electroformed Ni mesh with 320 $\mu$m grid spacing. Mesh cells are 
filled in a random pattern with 17 $\mu$m thick Ni, and the
whole mask is coated with 4 $\mu$m gold to provide $>$75\% absorption for 
E $<$ 20 keV. The overall
mask transmission is $\sim$40\%. The frame will also 
provide slots for $^{55}$Fe 
sources for in-flight detector calibration.
\begin{table}[ht]
\caption{BlackCAT Mask Parameters} 
\label{tab:mask}
\begin{center}       
\begin{tabular}{|l|l|} 
\hline
\rule[-1ex]{0pt}{3.5ex}  Focal length & 158 mm  \\
\hline
\rule[-1ex]{0pt}{3.5ex}  Mask size (aperture) & 170 $\times$ 88 mm   \\
\hline
\rule[-1ex]{0pt}{3.5ex}  Mask element & 320 $\mu$m $\times$ 320 $\mu$m   \\
\hline
\rule[-1ex]{0pt}{3.5ex}  Detector superpixel & 320 $\mu$m $\times$ 320 $\mu$m   \\
\hline
\rule[-1ex]{0pt}{3.5ex}  DM FOV & 0.95 sr   \\
\hline
\rule[-1ex]{0pt}{3.5ex}  Open element size & 263 $\mu$m $\times$ 263 $\mu$m  \\
\hline
\rule[-1ex]{0pt}{3.5ex}  Mask transmission & $>$40\%  \\
\hline
\rule[-1ex]{0pt}{3.5ex}  Image scale & 52"/pixel (6.9'/superpixel)  \\
\hline
\end{tabular}
\end{center}
\label{mask}
\end{table} 

{\bf Front-end processing board: }We plan to develop an FPGA-based Front-end Processing Board (FPB) to perform
high-speed digital processing of detector events including cosmic-ray rejection, event recognition, and event grading. The FPB receives data from the BlackCAT detectors and interfaces directly to the On-board Computer (OBC). The data stream will be stored in ring buffers on the FPB, sized to hold $\sim$2000 seconds worth of
data at the highest anticipated source plus background count rate during a bright GRB. Fast Fourier Transforms (FFTs) will be performed on the X-ray event data at superpixel (8$\times$8 pixel) resolution to form sky images. On the ground, further processing of the data will include back-projection at full detector resolution to obtain maximum angular resolution and sensitivity. This processing is very similar to what is done on the Swift BAT instrument.

\subsection{Spacecraft overview}
We will procure the BlackCAT spacecraft platform from Clyde Space, 
which has 
extensive heritage
developing standard CubeSat platforms and subsystems. Clyde's standard 6U incorporates all the
latest CubeSat technologies and offers a high level of reliability and system knowledge.
The payload is designed to fit within the available volume, mass, and interface constraints of the standard 6U
platform from Clyde and similarly, it has been designed to fit within the power and telemetry constraints imposed by standard power and telemetry systems that can fit within this 6U package.
The standard 6U package from Clyde space comes with Electrical Power 
Subsystem, Control and Data Handling board, Attitude Determination and Control System, and telemetry system.

The Electrical Power Subsystem (3G FlexU EPS) has multiple
regulated electrical buses (3.3, 5, 12, and 30 V) and an unregulated battery voltage bus ranging
from 6.2 V to 8.2 V. The EPS is designed to deliver $>90$W at maximum current. Electrical power is generated
by a Clyde Space 6U deployable solar panel system with a total of 18 UTJ large area cells populated in a 9s2p configuration. Primary power is routed from the array to the Solar Cell Assembly (SCA), 
where it is distributed (essential and switched) to the various loads and used for battery charging. The battery
system includes Lithium-Polymer batteries with 120 Whr energy storage capacity.

The Control And Data Handling (C\&DH) subsystem will be controlled by a Clyde Space High Speed On-board Computer (OBC) in the standard Clyde 6U platform. The OBC provides a highly integrated robust computing platform for space applications.

The telemetry subsystem will be provided by
partners of Clyde at French South African Institute of Technology (F'SATI). The telemetry system operates at 2.2-2.3 GHz band
and it is compatible with the CubeSat standard, with a CubeSat Kit PC/104 form factor. It implements QPSK modulation with transmission data rates up to 2 Mbps. A nadir-facing S-band
antenna of wide beam width will be incorporated into the CubeSat design providing ground station communication through a wide range of elevation angles. The CubeSat will also be capable of TDRSS Alert Messaging.

The ADCS consists of a star tracker, Sun sensor, temperature sensor,
3-axis reaction wheels system, magnetorquers, 3-axis rate gyros, and two 3-axis magnetic field sensors. 
The ADCS reads the sensor readings and processes the data using an Extended Kalman Filter to obtain an estimate of the satellite
attitude. This is followed up by a large set of propagation algorithms 
including calculating the coordinates of celestial bodies and the satellite's own state vector. The GPS provided with the OBC can
also be used to determine spacecraft attitude. The ADCS can be interfaced with the GPS in the OBC+GPS receiver bundle. These data are used to estimate the magnetic field experienced by the
satellite. With a known target attitude and estimated spacecraft attitude, a set of control algorithms
are then used to fine point to the target. The ADCS has the capability to re-tune all the algorithms while in orbit.
The ADCS is expected to provide pointing knowledge of $<$ 30 arcsec at 3$\sigma$ level and pointing accuracy better than 30 arcsec. The expected stability and jitter values are $\pm$0.0004 deg/s and $<$50 arcseconds, respectively. 

\subsection{Orbit and radiation environment}
BlackCAT will be proposed for a Low Earth Orbit (LEO) launch via NASA launch services. The
minimum orbital requirements are a near-circular, dawn-dusk, sun-synchronous orbit.
The altitude can be from 400 -- 600 km. 
This dawn-dusk orbit allows the continuous illumination of the solar panels while also providing
continuous observing. The sun synchronous orbit provides excellent ground coverage. We have
studied the radiation environment in this orbit and find that the Total Lifetime Dose the electronics are exposed to can be reduced to manageable levels with 3.5 mm thick aluminum shielding.
One of the primary reasons for our choice of hybrid CMOS detectors is the fact that they are radiation hard.

\subsection{Thermal design}
BlackCAT will point in anti-Sun direction. As a result, it has solar panels that are shading the spacecraft while
always pointing towards the Sun and has a large side of the 6U spacecraft that is always pointing
towards cold space. This provides an excellent position for the radiator.
The detector module (DM) employs active heater control for the 
detector mosaics and masks. Flight
Software (FSW) controls DM heaters to keep the detectors at -60°C $\pm$0.5°C, 
with $>$5°C margin in the thermal design. The electronics package is housed
in its own enclosure and dumps its
heat to the S/C structure. A finite
element thermal model of BlackCAT was created in Thermal Desktop® consisting of a mechanical
structure, avionics boxes with appropriate power output, detector
module with walls, zenith-pointing
radiator, and solar panels. The
model confirmed that the S/C
components could be arranged
such that only passive cooling is
required. 

\section{Project status and planning}\label{plan}
This conceptual design of the CubeSat mission has been submitted to NASA and is currently under review. Prototypes of the Speedster
detectors proposed for this payload have been tested in our laboratory. The full array of detectors will be available by mid-2019. We plan to start calibrating the detectors soon after their arrival and make them flight ready. If BlackCAT is selected for flight, we plan to first build and develop the coded mask and demonstrate the imaging technology in laboratory with the Speedster detectors. This will be followed up by testing of the S/C components independently in our laboratory, and finally completing end-to-end testing and calibration of the full payload.
 


\end{document}